\documentclass[twocolumn,prb,preprintnumbers]{revtex4}
\usepackage{natbib}
\usepackage{graphicx}
\usepackage{latexsym}
\usepackage{hyperref}
\usepackage[centertags]{amsmath}
\usepackage{amsfonts}
\usepackage{amssymb}
\usepackage{newlfont}
 \usepackage[latin1]{inputenc}
 \usepackage[T1]{fontenc}
\usepackage{latexsym}
\usepackage{epsfig}
\usepackage{makeidx}
\usepackage[dvips]{color}

\newcommand\peq{\footnotesize}

\begin{document}

\title{Kondo Effect and Spin Glass Behavior of Dilute Iron Clusters
in Silver Studied by M\"{o}ssbauer Spectroscopy and Resistivity}

\author{W. T. Herrera}\email[Electronic address: ]{william@cbpf.br}
\affiliation{Instituto de F\'{i}sica, Universidade do Estado de Rio
de Janeiro, Rua S\~{a}o Francisco Xavier 524, 20559-900, RJ, Brazil}

\author{S. M. Ramos}
\author{E. M. Baggio-Saitovitch} \affiliation{Centro
Brasileiro de Pesquisas F\'{i}sicas, Rua Dr. Xavier Sigaud 150,
22290-180, RJ, Brazil}

\author{F. J. Litterst}
\affiliation{Institute for Physics of Condensed Matter, Technische
Universit\"{a}t Braunschweig, Mendelssohnstrasse 3, 38106
Braunschweig, Germany}

\begin{abstract}
Thin films of silver containing $0.3-1.5$ $at$ $\%$ $Fe$ have been
prepared by vapor co-deposition. Depending on substrate temperature
and iron concentration we could systematically follow the formation
of nanometer size clusters of iron from initially dilute iron
monomers. samples were characterized via $X$-ray diffraction,
resistivity and M\"{o}ssbauer spectroscopic measurements.

The magnetic behavior derived from M\"{o}ssbauer data can be best
described with an ensemble of ferromagnetic mono-domain particles.
The magnetic freezing observed at low temperatures, is controlled
via the inter-particle interactions mediated via conduction electron
polarization,  i.e. $RKKY$ interaction.

The interaction of the cluster magnetic moments with the conduction
electron sea is best quantified by the electrical resistivity data.
For all studied concentrations we find a non-monotonic variation
with temperature which can be understood by competing shielding of
the cluster moments by conduction electron spin scattering due to
Kondo effect and the magnetic coupling.

\end{abstract}

\maketitle

\section{Introduction}
In the past $Fe$ precipitates in noble metals have been studied
following various tracks of preparation. Mostly precipitates are
formed from concentrated solid solutions under thermal treatment.
Less common is starting out from very dilute Fe impurities embedded
in the metal host and subsequent annealing leading to diffusion of
impurities and clustering. For characterizing the electronic state
of Fe in these precipitates and its magnetic properties
M\"{o}ssbauer spectroscopy is a highly suitable method.

Of particular interest are systems with very low miscibility in
solid and liquid state, like e.g. $Ag$:$Fe$. Even very dilute alloys
with low concentration of $Fe$ in $Ag$ may only be achieved via
non-equilibrium preparation, i.e. vapor
deposition\cite{morales1,larica,elisa}, electron-beam
co-evaporation\cite{peng}, mechanical alloying\cite{gomez} or
implantation\cite{longworth,marest}. Well known are the early
M\"{o}ssbauer spectroscopic studies of ppm concentrations of $Fe$ in
$Ag$ prepared via diffusion of $^{57}Co$ into the host matrix with
the radioactive decay of $^{57}Co$ leading to
$^{57}Fe$\cite{steiner}.

Morales et al \cite{morales1} succeeded in preparing films in the
range of percent concentration of $Fe$ in $Ag$ by evaporating the
elements in proper proportion and co-depositing them onto Kapton
substrates kept at 16 $K$. From the M\"{o}ssbauer spectra isolated
monomeric and dimeric $Fe$ in the $Ag$ matrix could be clearly
identified as major components. In addition minor contribution was
found which was attributed to clusters of fcc iron.

Upon annealing at room temperature the formation of clusters is enhanced, but only
for annealing around 480 $K$ bcc $Fe$ precipitates are formed. An
identification of the various species is possible from their
distinctly different hyperfine parameters.

\section{Experimental}
In continuation of these studies we prepared films by co-deposition
of $Fe$ and $Ag$ with nominal concentrations between 0.3 and 1.5
$at$ $\%$ of $Fe$. We have characterized the samples by X-ray
diffraction, $^{57}Fe$ M\"{o}ssbauer spectroscopy and electrical
resistivity measurements. In order to receive a sufficient amount of
$^{57}Fe$ for performing M\"{o}ssbauer experiments we used iron
metal enriched to $90$\% of $^{57}Fe$. The preparation procedure and
the evaporation facility were the same as described in refs.
\cite{morales1,larica,elisa}. The base vacuum pressure in the
deposition chamber was $2\times10^{-9}$ $mbar$, increasing to
$2\times 10^{-8}$ $mbar$ during deposition. The films were deposited
onto Kapton foils kept either at 285 $K$ or 85 $K$. The deposition
rate was monitored using piezo-crystals and typical values were 4
{\AA}$/s$ for $Ag$ and 0.02 \AA$/s$ for $Fe$ (see table
\ref{tabla1}). The total film thicknesses were typically 1000 to
4000 $nm$. Under these preparation conditions we expected clusters
to be formed directly during the deposition process and not only
after annealing.

First sets of M\"{o}ssbauer experiments were performed in situ, i.e.
in the cryostat where the preparation was done. The spectrometer was
of standard type with sinusoidal velocity sweep. The $^{57}CoRh$
source was kept at room temperature. It turned out that the films
could be transferred to a variable temperature cryostat ($1.5K -
300K$) without observing any change of the spectra. So it was
possible to perform ex situ experiments without inducing structural
changes of the film.

Resistivity measurements were taken between 1.5 $K$ and 300 $K$
using a standard 4 contacts technique. X-ray diffraction was
performed on a Rigaku MiniFlex using $Cu$ $K_{\alpha 1}$ and
$K_{\alpha 2}$.

{\renewcommand{\arrayrulewidth}{1.2pt}
\begin{table}[h!]
\renewcommand{\arraystretch}{1.3}\renewcommand{\tabcolsep}{0.16cm}
\begin{tabular}{@{\hspace{0.0cm}}p{1.2cm} c c c c c@{\hspace{0.0cm}}}
 \hline\hline
  \peq{$at$ \% $Fe$}& \peq{$\eta_{Fe}$ ({\AA}$/s$)} & \peq{$\eta_{Ag}$ ({\AA}$/s$)}&
  \peq{$T_s(K)$} & \peq{$t$ ($min$)} & \peq{$\xi$ ($nm$)}\\
  \hline
  0.3 & 0.009 & 4.0  & 285 & 120 & $\approx$ 2886 \\
  0.5 & 0.01  & 3.0  & 285 & 115 & $\approx$ 2076 \\
  0.8 & 0.007 & 1.3  & 285 & 124 & $\approx$ 972  \\
  1.0 & 0.028 & 4.05 & 283 & 117 & $\approx$ 2862 \\
  1.2 & 0.031 & 1.55 & 283 & 120 & $\approx$ 1138 \\
  1.5 & 0.035 & 4.6  & 280 & 130 & $\approx$ 3615 \\
  \hline
  1.0 & 0.02  & 4.5  & 85  & 124 & $\approx$ 3349 \\
  0.4 & 0.01  & 3.9  & 85  & 140 & $\approx$ 3307 \\
  \hline
  0.0 & 0.0   & 3.0  & 285 & 140 & $\approx$ 2322 \\
  \hline\hline
\end{tabular}
%Upsilon
\caption{\label{tabla1} \peq{$\eta$ deposition rate,  $T_s$
substrate temperature, $t$ deposition time and $\xi$ thickness.}}
\end{table}}

\section{Results and discussion}
\begin{figure}%[h!]
 \centering\includegraphics[scale=0.83]{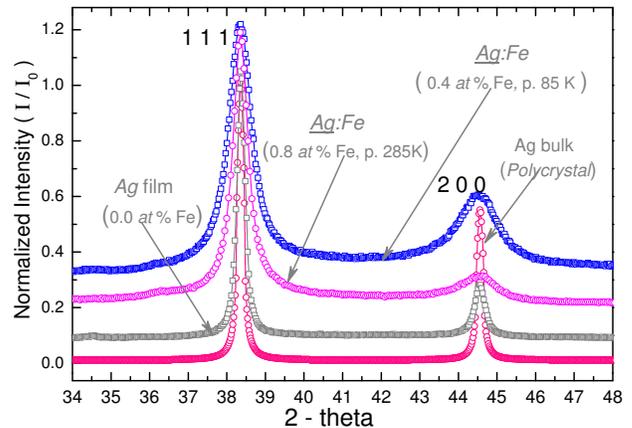}
\caption{\label{nano1} \peq{X-ray diffraction  of polycrystalline
Ag, a pure Ag film and iron doped Ag films deposited onto 285 $K$
and 85 $K$ substrates.}}
\end{figure}

X-ray diffraction of the iron doped films revealed significant
broadenings of the silver Bragg peaks indicating the formation of
small grains. Using the Scherrer formula we estimate mean grain
sizes of about 22 $nm$ and 11 $nm$ for films deposited onto 285 $K$
and 85 $K$ substrates, respectively. All films reveal a high degree
of texture (see fig. \ref{nano1}).

\begin{figure}%[h!]
 \centering\includegraphics[scale=1.78]{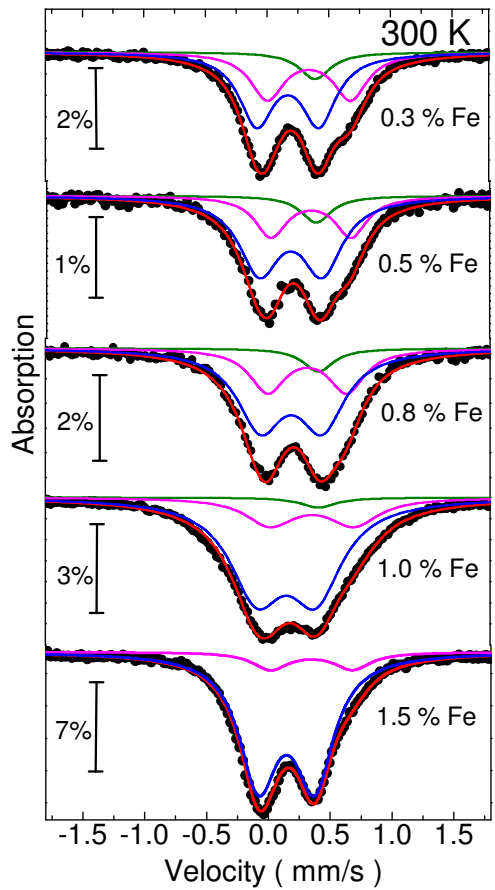}
\caption{\label{todos1} \peq{Mössbauer spectra taken at 300 $K$ for
0.3, 0.5, 0.8, 1.0 and 1.5 $at$ $\%$ $Fe$.}}
\end{figure}

Fig. \ref{todos1} it shows a series of M\"{o}ssbauer spectra taken
at 300 $K$ for various concentrations of iron in films prepared at
285 $K$. Below 1 $at$ $\%$  $Fe$ the spectra are composed by a
superposition of three subspectra. One singlet line $S$ can be
identified as being caused by iron
monomers\cite{morales1,longworth,kataoka1,pereira}; we observe in
addition two quadrupole split doublets which for all concentrations
have practically the same hyperfine parameters. In contrast to the
earlier samples prepared by vapor quenching onto substrates kept at
$16 K$ \cite{morales1}, we find no resolved contribution from iron
dimers. There is also no indication for the presence of fcc clusters
of the type that was found after deposition at low temperatures. We
therefore performed a re-analysis of these earlier data and it
turned out that they are even better adjusted by the monomer, the
dimer and the same two doublet contributions as found for the new
samples.

\begin{figure}%[h!]
 \centering\includegraphics[scale=1.3]{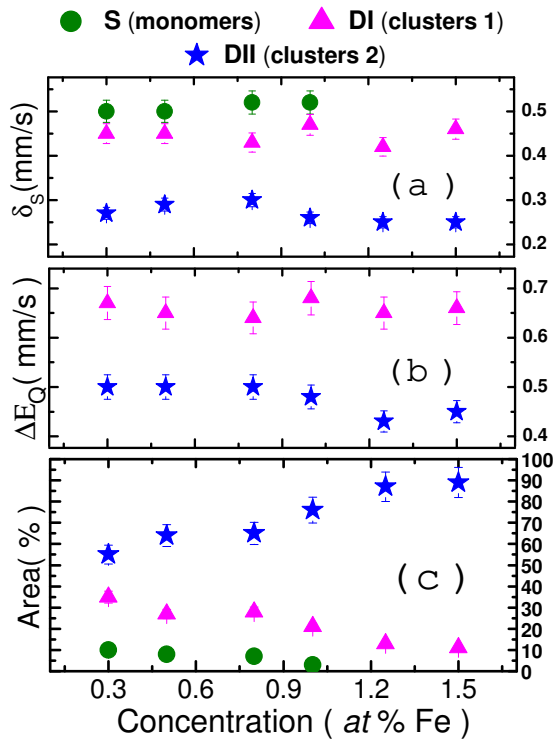}
\caption{\label{para1} \peq{Hyperfine parameters. (a) isomer shift
$\delta_S$, (b) quadrupole splitting $\Delta E_Q$, and (c) relative
spectral areas of the subspectra S (monomers), DI and DII
(clusters).}}
\end{figure}

We label these two doublets with $DI$ and $DII$. The hyperfine
parameters and the variation of their relative intensities for the
three species are shown in figure \ref{para1}. The relative amount
of monomers is continuously decreasing with increasing iron
concentration and finally for more than 1 $at$ $\%$  $Fe$ the
singlet sub-spectrum can be no more traced and only a superposition
of the two quadrupole doublets is found ( fig. \ref{todos1} ).


Whereas the relative spectral area of $DI$ is decreasing with
increasing iron concentration, the area of $DII$ is increasing. A
close lying possible interpretation for the two doublets would be to
attribute them to iron on the surface and from the core of one type
of cluster. From the variation of relative areas with concentration
one then would have to relate $DI$ to the surface and $DII$ to the
core which however would imply cluster sizes in the micrometer
range. This definitively can be excluded since such clusters should
have fcc or bcc structure and reveal magnetic blocking in the 100
$K$ range or even higher what is not observed (see below).

We attribute the two doublets to clusters in the nanometer range,
with $DI$ representing smaller and $DII$ bigger clusters. $DI$ has
the higher isomer shift rather close to the value of the monomer and
a big quadrupole splitting indicative for a low symmetry
surrounding. The isomer shift for $DII$ is smaller meaning that the
number of iron neighbors is higher; the quadrupole splitting is
smaller, i.e. coming closer to a cubic arrangement. Such clusters
are supposed not to fit into the fcc lattice of silver and should
appear at grain boundaries especially in view of the small grain
structure of our films.
\begin{center}
\begin{figure}%[h!]
 \centering\includegraphics[scale=0.38]{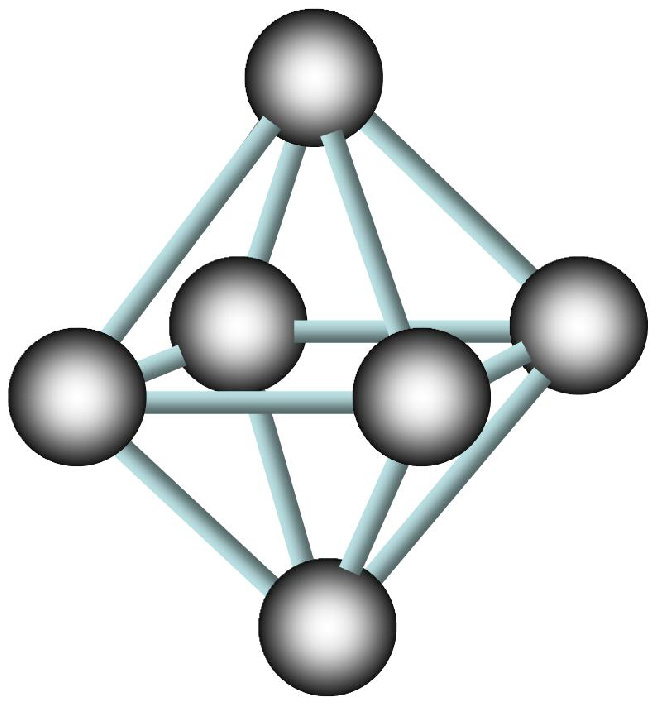}\includegraphics[scale=0.52]{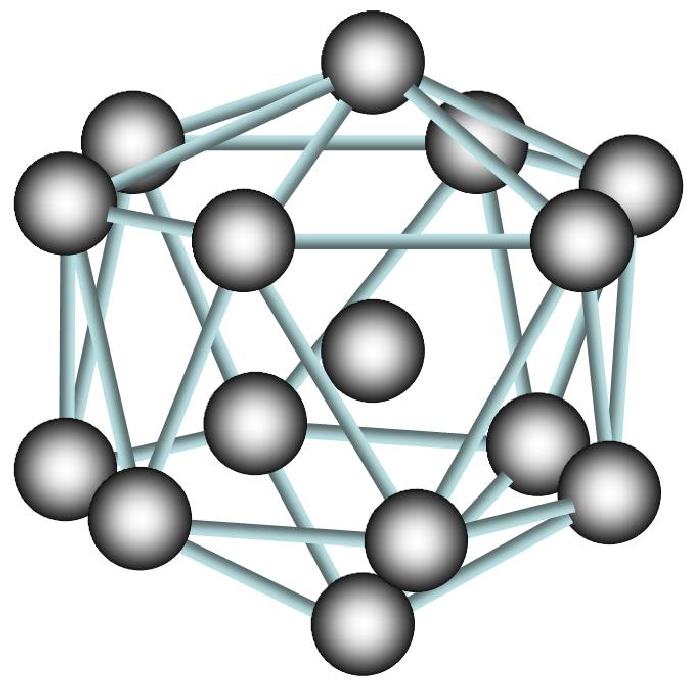}
\caption{\label{clusters}\peq{Clusters with 6 and 15 atoms.}}
\end{figure}
\end{center}

As tentative model one could consider some stable free iron clusters
as proposed from theoretical calculations by Rollman \textit{et
al}\cite{Rollmann}, e.g. clusters $Fe_6$ and $Fe_{15}$ (to see fig.
\ref{clusters}). Note that these clusters are so small that
practically all iron atoms are on surfaces and the spectral
contribution from the core of the particles is zero or negligible.

\begin{figure}%[h!]
 \centering\includegraphics[scale=1.28]{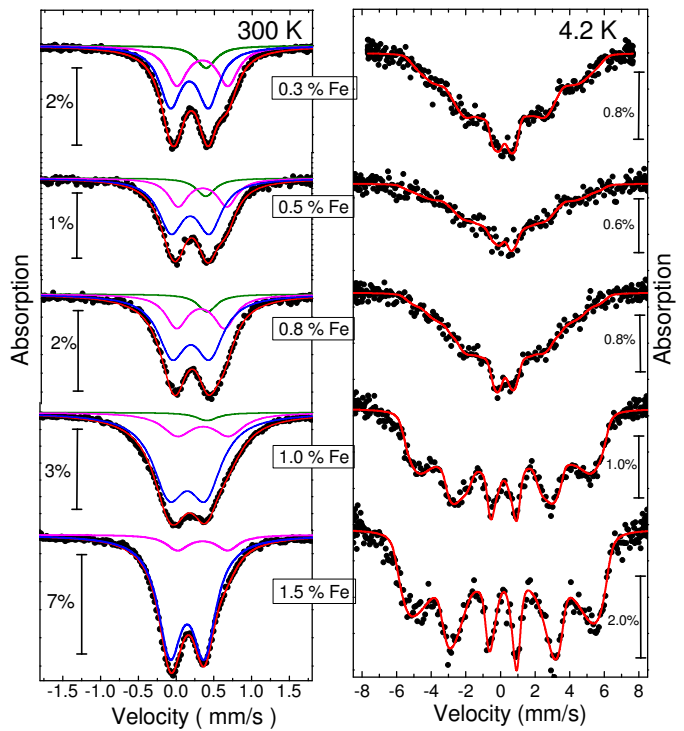}
\caption{\label{todos2} \peq{Mössbauer spectra taken at 300 $K$ and
4.2 $K$ for 0.3, 0.5, 0.8, 1.0 and 1.5 $at$ $\%$ $Fe$.}}
\end{figure}

For temperatures below about 15 $K$ we find an onset of magnetic
hyperfine interaction connected with the freezing of the magnetic
moments of the clusters. The spectra taken at 4.2 $K$ are shown in
fig. \ref{todos2}.
It is clearly seen that there is a distribution of the magnetic
hyperfine interaction which becomes better defined with increasing
iron concentration, i.e. with larger number of clusters. The drawn
lines through the data points of the spectra are simplistic fits
using a distribution of hyperfine fields. For the present discussion
it may suffice to notice the onset of magnetic interaction. A more
detailed interpretation of these patterns (e.g. taking into account
the role played by magnetic fluctuations, cluster-cluster
interactions etc.) will be presented elsewhere together with
susceptibility, magnetization and M\"{o}ssbauer measurements in
applied magnetic field over a wide temperature range. These data
yield further support for small clusters with moments on the order
of about $20-50$ $\mu_B$.

\begin{figure}%[h!]
 \centering\includegraphics[scale=0.82]{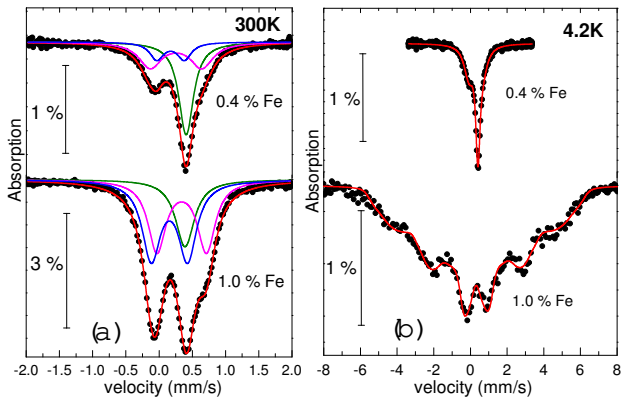}
\caption{\label{todos85K} \peq{Mössbauer spectra taken at 300 $K$
and 4.2 $K$ for 0.4 and 1.0 $at$ $\%$ $Fe$ (samples prepared at
85K).}}
\end{figure}

The room temperature spectra of the samples prepared at 85 $K$
reveal the same three species $S$, $DI$ and $DII$, yet with a
clearly higher contribution by the monomer $S$ (see fig.
\ref{todos85K}). At 4.2 $K$ the sample with 1.0 $at$ \%  $Fe$ shows
a magnetic hyperfine pattern which however indicates a lower degree
of freezing than found for the sample prepared at 285 $K$. For the
sample with 0.4 $at$ \%  $Fe$  there is still no indication of a
magnetic hyperfine interaction (fig. \ref{todos85K}) at 4.2 $K$.

When comparing the freezing behaviour one has to take into account
that the concentration of clusters in the samples prepared at 295
$K$ (with dominant bigger clusters) is in fact lower than in the
samples prepared at 85 $K$, so their distances are larger than those
between the isolated monomers for the same nominal concentration of
iron. Their interaction however, is increased due to the augmented
moment of each cluster and one can thus understand the stronger
tendency for freezing with increasing concentration despite longer
distances between clusters. Using a RKKY model adopted for
clusters\cite{Altbir} with the known Fermi surface vectors of silver
and ferromagnetic cluster moments with about 40 $\mu_B$ we arrive
for our concentrations at average cluster-cluster interaction
energies corresponding to $10-20$ $K$. This is in agreement with the
freezing temperatures derived from the Mössbauer data.

\begin{figure}%[h!]
 \centering\includegraphics[scale=1.3]{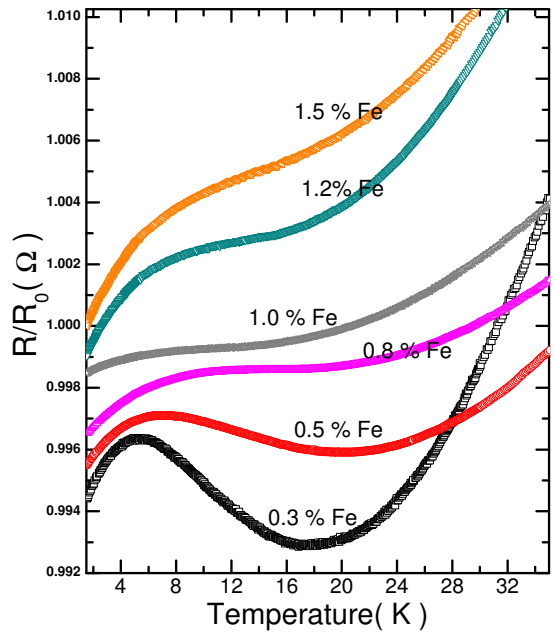}
\caption{\label{res1} \peq{ Electrical resistivity for 0.3, 0.5,
0.8, 1.0, 1.2 and 1.5 $at$ \%  $Fe$ prepared at 285K, normalized to
$R_0$ ($T=1.5K$)}}
\end{figure}

\begin{figure}%[h!]
 \centering\includegraphics[scale=1.2]{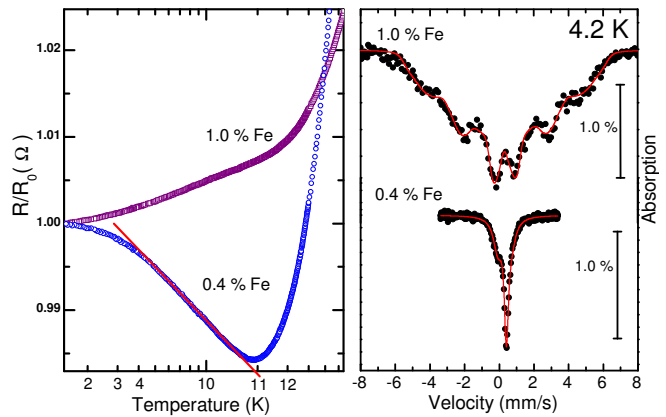}
 \caption{\label{res2} \peq{Electrical resistivity for 0.4 and 1.0 $at$ \%  $Fe$
 prepared at 85K, normalized to $R_0$ ($T=1.5K$).}}
\end{figure}

The resistivity results for the samples prepared at 285$K$ and 85$K$
are shown in the Fig. \ref{res1} and Fig. \ref{res2}, respectively.
For low iron concentrations we find clear minima in the temperature
dependence of the resistivity (typically around 20 $K$) followed by
maxima at lower temperatures. The minima are indicative for the
onset of Kondo effect. The typical logarithmic increase of
resistivity with decreasing temperature may most clearly be traced
for the samples with low concentrations, i.e. with a high number of
monomers. Saturation is only achieved below about 2 $K$ which is in
agreement with earlier data obtained by Mössbauer\cite{steiner},
TDPAC \cite{riegel1,riegel2}  and susceptibility \cite{hanson}
measurements on $Fe$ monomers in bulk silver where Kondo
temperatures of about $1.5 - 2$ $K$ were found.

Actually an increased value was derived from TDPAC \cite{mishra} for
dilute iron impurities in nanocrystalline silver. This discrepancy
of $T_K$ found between bulk and nanocrystalline silver was
attributed to a pressure induced increased hybridization of iron
$3d$ electrons with the conduction electrons near the grain surfaces
\cite{crepieux,crepieux2}. The silver grain size of 19 $nm$ is
actually nearly the same as found for our samples prepared at 285
$K$ with grain sizes of about 22 $nm$, however, our data do not
support an increase of Kondo temperature.

\begin{figure}%[h!]
 \centering\includegraphics[scale=0.7]{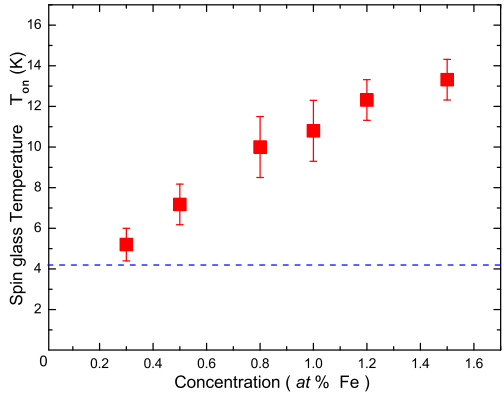}
 \caption{\label{para2} \peq{Onset temperature $T_{on}$ for spin freezing derived from resistivity data.}}
\end{figure}

The mentioned resistivity maxima are observed for samples containing
a higher amount of clusters and also revealing spin freezing from
M\"{o}ssbauer effect. The decrease of resistivity for temperatures
below the maxima is related to the onset of the magnetic interaction
between the clusters. It is getting clearly stronger with increasing
concentration of iron, i.e. also with number of clusters. The shape
of the resistivity curves are in very good qualitative agreement
with the calculations of Larsen \cite{larsen} and  more recently by
Vavilov et al \cite{vavilov} predicting  a non monotonic variation
of resistivity with temperature when in concentrated systems the
RKKY interaction between magnetic impurities is strong compared to
the Kondo interaction. Similar observations have been reported,
e.g., for $Au$ and $Cu$ matrices doped with $Mn$ and $Fe$
impurities, there however at much lower
concentrations\cite{schilling}. For concentrations above about 1
$at$ $\%$ $Fe$ the resistivity minimum can no more clearly be traced
in our data and only the turn-down of resistivity due to the
magnetic freezing is visible. Despite the temperature where the
maximum of resistivity occurs does not directly correspond to the
spin glass temperatures as derived e.g. from the susceptibility cusp
(see refs. \cite{larsen,schilling}), we plot this temperature
$T_{on}$ in fig. \ref{para2} for demonstrating the concentration
dependence of spin freezing. The values were obtained from the first
derivative for 0.3, 0.5 and 0.8 $at$ \% $Fe$ and from the second
derivate for 1.0, 1.2 and 1.5 $at$ \%. A straight line in fig.
\ref{para2} indicates the $T=4.2$ $K$ the temperature at which the
M\"{o}ssbauer measurements were taken. This illustrates why the
spectra for $Fe$ $<$ 1.0 $at$ \% are not static since the freezing
is very close to 4.2 $K$.

For the higher concentrations we take the temperatures where the
downturn of resistivity from the phonon contribution becomes
apparent. These values between about 5 and 15 K are again in fair
agreement with the observed onset of magnetic hyperfine interaction
from the M\"{o}ssbauer data for these concentrations. The
disappearance of the Kondo minimum occurs in parallel to the
vanishing of the monomer contribution in the Mössbauer spectra and
also of the clusters of type $DI$.  There is no clear indication for
Kondo scattering when clusters of type $DII$ become dominant. We
therefore have to conclude that the Kondo anomalies observed in the
resistivity data are mainly caused by contributions from monomeric
iron and eventually of clusters $DI$. In any case the Kondo
temperatures for samples with iron concentrations above about 1 $at$
\% must to be below about 3 $K$.
\section{Conclusions}
In summary we have observed the formation of well defined nano-sized
iron clusters in silver films prepared by co-deposition from iron
and silver atomic beams under varying conditions and for a series of
concentrations. Resistivity data give clear evidence for Kondo
effect for samples containing monomeric iron impurities and clusters
of type $DI$ but not for clusters of type $DII$, i.e for iron
concentrations up to about 1 $at$ \%. Samples containing clusters
reveal the typical non-monotonic resistivity behaviour expected for
spin glass freezing competing with Kondo effect. The spin glass
freezing is clearly visible from the appearance of magnetic
hyperfine splitting. The spin glass transition temperature is in
good agreement with estimates from a RKKY model for interacting
clusters.
\section*{ACKNOWLEDGMENTS}
%\begin{acknowledgements}
Financial support from FAPERJ, CNPQ, CAPEs and DAAD are acknowledged
by the authors.
%\end{acknowledgements}

\end{document}